\documentclass[a4paper,aps,pra,twocolumn,preprintnumbers,amsmath,amssymb,superscriptaddress]{
revtex4}
\usepackage{amsmath}
\usepackage{verbatim}
\usepackage{graphicx}

\usepackage{color}

 \newcommand{\ket}[1]{\left|#1\right\rangle} 
 \newcommand{\bra}[1]{\left\langle#1\right|} 

\begin{document}
 
\title{Quantum annealing for the number partitioning problem using a tunable spin glass of ions}

\author{Tobias Gra{\ss}}
\email{tobias.grass@icfo.es}
\affiliation{ICFO-Institut de Ci\`encies Fot\`oniques, The Barcelona Institute of Science and Technology, Av. Carl-Friedrich Gauss 2, 08860 Castelldefels, Spain}

\author{David Ravent\'os}
\affiliation{ICFO-Institut de Ci\`encies Fot\`oniques, The Barcelona Institute of Science and Technology, Av. Carl-Friedrich Gauss 2, 08860 Castelldefels, Spain}

\author{Bruno Juli\'a-D\'iaz}
\affiliation{ICFO-Institut de Ci\`encies Fot\`oniques, The Barcelona Institute of Science and Technology, Av. Carl-Friedrich Gauss 2, 08860 Castelldefels, Spain}
\affiliation{Departament de F\'isica Qu\`antica i Astrof\'isica,\\
Facultat de F\'{\i}sica, Universitat de Barcelona, 08028 Barcelona, Spain}
\affiliation{Institut de Ci\`encies del Cosmos, Universitat de Barcelona, ICCUB, Mart\'i i
Franqu\`es 1, 08028 Barcelona, Spain}

\author{Christian Gogolin}
\affiliation{ICFO-Institut de Ci\`encies Fot\`oniques, The Barcelona Institute of Science and Technology, Av. Carl-Friedrich Gauss 2, 08860 Castelldefels, Spain}
\affiliation{Max-Planck-Institut f{\"u}r Quantenoptik, Hans-Kopfermann-Stra{\ss}e 1, 85748 Garching, Germany}

\author{Maciej Lewenstein}
\affiliation{ICFO-Institut de Ci\`encies Fot\`oniques, The Barcelona Institute of Science and Technology, Av. Carl-Friedrich Gauss 2, 08860 Castelldefels, Spain}
\affiliation{ICREA-Instituci\'o Catalana de Recerca i 
Estudis Avan\c{c}ats, Llu\'is Companys 23, 08010 Barcelona, Spain}

\maketitle

{\bf 
Exploiting quantum properties to outperform classical ways of information-processing is an outstanding goal of modern physics. A promising route is quantum simulation, which aims at implementing relevant and computationally hard problems in controllable quantum systems. Here we demonstrate that in a trapped ion setup, with present day technology, it is possible to realize a spin model of the Mattis type that exhibits spin glass phases. Remarkably, our method produces the glassy behavior without the need for any disorder potential, just by controlling the detuning of the spin-phonon coupling. Applying a transverse field, the system can be used to benchmark quantum annealing strategies which aim at reaching the ground state of the spin glass starting from the paramagnetic phase. In the vicinity of a phonon resonance, the problem maps onto number partitioning, and instances which are difficult to address classically can be implemented.}

\vspace{1cm}

Spin models are paradigms of multidisciplinary science: They find several applications
in various fields of physics, from condensed matter to high energy 
physics, but also beyond the physical sciences. In neuroscience, the famous Hopfield model describes brain functions such as associative memory by an interacting spin system \cite{hopfield}.
This directly relates to computer and information sciences, where pattern recognition or error-free coding can be achieved using spin models~\cite{nishimori}. 
Importantly, many optimization problems, like number partitioning or the travelling salesman problem, belonging to the class of NP-hard problems, can be 
mapped onto the problem of finding the ground state of a specific spin 
model~\cite{barahona,lucas}. This implies that solving spin models is a 
task for which no general efficient classical algorithm is known to exist.
In physics, analytic replica methods have been developed in the context of spin glasses~\cite{sk,parisi}. A controversial development, 
supposed to provide also an exact numerical understanding of spin glasses, 
regards the D-Wave machine. Recently introduced on the market, this device solves classical spin glass models, but the underlying mechanisms are not clear, and it remains an open question whether it provides a speed-up over the best classical algorithms~\cite{troyer2014,katzgraber,google}. 

This triggers interest in alternative quantum systems designed to solve general spin models via quantum simulation. A noteworthy physical system for this goal are trapped ions: Nowadays, spin systems of trapped ions are available in many laboratories~\cite{schaetz-natphys,monroe-spinspin,kim2010,britton2012,jurcevic,richerme}, and adiabatic state preparation, similar to quantum annealing, is experimental state-of-art. Moreover, such system can exhibit long-range spin-spin interactions~\cite{porras04} mediated by phonon modes, leading to a highly connected spin model. Here we demonstrate how to profit from these properties, using trapped ions as a quantum annealer of a classical spin glass model. 

We consider a setup described by a time-dependent Dicke-like model:
\begin{align}
\label{H0}
H_0(t)  =& \sum_{m} \hbar \omega_m a^\dagger_m a_m + \sum_{i,m} \hbar \Omega_i \sqrt{\frac{\omega_{\mathrm{recoil}}}{\omega_m}} \xi_{im}  \sin( \omega_{\rm L} t )\nonumber \\ 
&  \times \sigma_x^i (a_m + a_m^\dagger),
\end{align}
with $a_m$ annihilating a phonon in mode $m$ with frequency $\omega_m$ and characterized by the normalized collective coordinates $\xi_{im}$. The second term in $H_0$ couples the motion of the ions to an internal degree of freedom (spin) through a Raman beam which induces a spin flip on site $i$, described by $\sigma_x^i$, and (de)excites a phonon in mode $m$. Here, $\Omega_i$ is the Rabi frequency, $\hbar \omega_{\rm recoil}$ the recoil energy, and $\omega_{\rm L}$ the beatnote frequency of the Raman lasers. Before also studying the full model, we consider a much simpler effective Hamiltonian, derived from Eq. (\ref{H0}) by integrating out the phonons \cite{mintert,porras04,schmitz,freericks}. The model then reduces to a time-independent Ising-type spin Hamiltonian
\begin{align}
\label{HJ}
 H_J = - \hbar \sum_{ij} J_{ij}\, \sigma_x^i \, \sigma_x^j.
\end{align}
Each phonon mode contributes to the effective coupling $J_{ij}$ in a factorizable way, proportional to $\xi_{im}\xi_{jm}$, and weighted by the inverse of the detuning from the mode $\delta_m = \omega_m- \omega_{\mathrm{L}}$ :
\begin{align}
\label{Jij}
J_{ij} = \Omega_i \Omega_j \frac{\omega_{\mathrm{recoil}}}{2\omega_{\rm L}} \sum_m \frac{\xi_{im} \, \xi_{jm}}{\delta_m}.
\end{align}
The $\xi_{im}$ imprint a pattern to the spin configuration, similar to the associative memory in a neural network~\cite{hopfield,sibylle}. The connection between multi-mode Dicke models with random couplings, the Hopfield model, and spin glass physics has been the subject of recent research~\cite{sachdev2011,lev2011,rotondo}, and the possibility of addressing number partitioning was mentioned in this context~\cite{rotondo}.

Before proceeding, we remind the reader that the concept of a spin glass used in the literature may have different and controversial meanings (cf. \cite{Stein,Stein-book,Moore,Katzgraber1}): i)  long-range spin glass models \cite{sk} and neural networks \cite{amitPRL,Amit1}, believed to be captured by the Parisi picture \cite{Parisi-true,Mezard-book} and replica symmetry breaking. These lead to hierarchical organization of the exponentially degenerated free energy landscape, breakdown of ergodicity and aging, slow dynamics due to a continuous splitting of the metastable states with decreasing temperatures (cf. \cite{Vincent}). ii) Short-range spin glass models believed to be captured by the Fisher-Huse \cite{Fisher} droplet/scaling model with a single ground state (up to a total spin flip), but a complex structure of the domain walls. For these models, aging, rejuvenation and memory, if any, have different nature and occurrence \cite{Vincent, Lefloch}; iii) Mattis glasses 
\cite{mattis}, where the huge ground state degeneracy becomes an exponential quasi-degeneracy, for which finding the ground state becomes computationally hard (\cite{mertens98}, see Subsection ``Increasing complexity''). Note that exponential (quasi)degeneracy of the ground states (or the free energy minima) characterizes also other interesting states: certain kinds of spin liquids or spin ice, etc.  

Here we analyse the trapped ion setup. Even without explicit randomness ($\Omega_i=\Omega={\mathrm{const.}}$), the coupling to a large number of phonon modes 
suggests the presence of glassy behavior. This intuition comes from the fact that the associative memory of the related Hopfield model works correctly if the number of patterns is at most $0.138 N$, 
with $N$ the number of spins~\cite{amitPRL}. For a larger number of patterns, the Hopfield model exhibits glassy behavior since many patterns have similar energy and the dynamics gets stuck in local minima. However, it is not clear~\textit{a priori} how the weighting of each pattern, present in Eq.~(\ref{Jij}), modifies the behavior of the spin model. In certain regimes the detuning suggests to neglect the contributions from all but one mode, leading to a Mattis-type model with factorizable couplings~\cite{mattis}, $J_{ij} \propto \xi_{im}\xi_{jm}$. Strikingly, 
the possibility of negative detuning, i.e., antiferromagnetic coupling to a pattern, drives such system into a glassy phase, characterized by a huge low-energy Hilbert space. The antiferromagnetic Mattis model is directly connected to the potentially NP-hard task of number partitioning~\cite{mertens98,Mertens2003}. Its solution can then be found via quantum annealing, i.e. via adiabatic ramping down of a transverse magnetic field, see also the proposal of Ref.~\cite{porras} for a frustrated ion chain.

We start by giving analytical arguments to demonstrate glassy behavior in the classical spin chains. Using exact numerics, we then focus on the quantum Mattis model. By calculating the magnetic susceptibility, and an Edward-Anderson-like order parameter, we distinguish between glassy, paramagnetic, and ferromagnetic regimes. The annealing dynamics is investigated using exact numerics and a semi-classical approximation. We demonstrate the feasibility of annealing for glassy instances. Finally, we show that the memory in the quantum Hopfield model is real-valued rather than binary. This property might be useful for various applications of quantum neural networks such as pattern recognition schemes.

 \section*{\large Results}
 
 \begin{figure}[t]
\centering
\includegraphics[width=0.47\textwidth, angle=0]{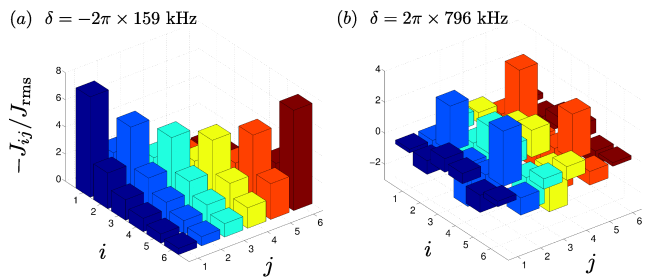}
\caption{\label{fig1}
{\bf Power-law decay vs. quasi-random patterns.} We plot the coupling constants $J_{ij}$ (in units of $J_{\rm rms} \equiv \frac{1}{N(N-1)}\sum_{i \neq j} |J_{ij}|^2$), for a system of six ions at different detunings $\delta = \omega_N - \omega_{\mathrm{L}}$. Rabi frequencies are taken as constants $\Omega_i={\rm const.}$ {\bf (a)} For negative detuning, $\delta= 2\pi \times 159$ kHz, interactions have a power-law decay. {\bf (b)} For positive detuning, $\delta = 2\pi \times 796$ kHz, the coupling constants resemble a spin glass.}
\end{figure}

\subsection*{Tunable spin-spin interactions}

We start our analysis by inspecting the phonon modes. For a sufficiently anisotropic trapping potential, the ions self-arrange in a line along the $z$-axis \cite{steane,james}. The phonon Hamiltonian $H_{\mathrm{ph}}$ is obtained by a second-order expansion of Coulomb and trap potentials around these equilibrium positions $z_i$: 
$H_{\mathrm{ph}} = (m/2)\sum_{ij} V_{ij} q_i q_j$, with $q_i$ the displacement of the $i$th ion from equilibrium. For the transverse phonon branch, $V_{ij}$ is given by~\cite{porras04}
\begin{eqnarray}
V_{ij} 
&= \left\{ \begin{array}{ll} \omega^2_{\mathrm{\perp}} - \frac{e^2/m}{4\pi \epsilon_0} \sum_{i'' (\neq i)}
\frac{1}{|z_i-z_{i''}|^3}, \hspace{0.5cm} & i=j
                 \\ \frac{e^2/m}{4\pi \epsilon_0} \frac{1}{|z_i-z_j|^3}. & i \neq j \end{array}
\right.
\end{eqnarray}
Our exact numerical simulations have been performed for six $^{40}$Ca$^+$ ions in a trap of frequency $\omega_{\mathrm{\perp}}=2\pi\times2.655$ MHz, as used in a recent experiment \cite{jurcevic}. In order to maximize the bandwidth of the phonon spectrum and thereby facilitate the annealing process, we choose the radial trap frequency $\omega_z$ as large as allowed to avoid zig-zag transitions, $\omega_z \lesssim 1.37 \omega_\perp \cdot N^{-0.86}$. 
Diagonalizing the matrix $V_{ij}$ leads to the previously introduced mode vectors ${\boldsymbol\xi}_m=(\xi_{1m},\dots,\xi_{Nm})$, which are normalized to one, and ordered according to their frequency $\omega_m$:
\begin{align}
 {\boldsymbol\xi}_{m'}^T V {\boldsymbol\xi}_m = \omega_{m}^2 \delta_{m,m'}.
\end{align}
The mode $\boldsymbol\xi_N$ with largest frequency, $\omega_N=\omega_{\mathrm{\perp}}$, is the center-of-mass 
mode, $\xi_{iN}=N^{-1/2}$. Parity symmetry of $V_{ij}$ is reflected by the modes, $\xi_{im} = \pm \xi_{(N+1-i)m}$, and even ($+$) and odd ($-$) modes alternate in the phonon spectrum. We focus on even $N$, for which all components $\xi_{im}$ are non-zero. Except for the center-of-mass mode, all modes fulfill $\sum_i \xi_{im} = 0$. 

Previous experiments \cite{schaetz-natphys,monroe-spinspin,kim2010,britton2012,jurcevic} have mostly been performed with a beatnote frequency $\omega_{\mathrm{L}}$ 
several kHz above $\omega_N$, leading to an antiferromagnetic coupling $J_{ij} <0$ 
with power-law decay, see Fig.~\ref{fig1}(a). Despite 
the presence of many modes, the couplings $J_{ij}$ then take an ordered structure. 
This work, in contrast, focuses on the regime $\omega_{\mathrm{L}} < \omega_N$, 
where modes with both positive and negative $\xi_{im}$ contribute, and ferro- and antiferromagnetic couplings coexist, cf. Fig.~\ref{fig1}(b). 
This reminds of the disordered scenario of common spin glass models like the 
Sherrington-Kirkpatrick model~\cite{sk}. In the following we study the properties of the time-independent effective spin model, before considering the full time-depedent problem involving spin-phonon coupling.

\subsection*{Classical Mattis model}

Close to a resonance with one phonon mode, simple 
arguments allow for deducing the spin configurations of the ground states. In 
this limit, we can neglect the other modes, leading to a Mattis model. A single pattern $\boldsymbol\xi_m$ then determines the coupling, $J_{ij} \propto \xi_{im} \xi_{jm}$. 
The sign of $J_{ij}$ depends on the sign of the detuning: Below the resonance, 
we have ${\mathrm{sign}}(J_{ij}) = {\mathrm{sign}}(\xi_{im}\xi_{jm})$, and accordingly the 
energy $-\hbar J_{ij}\sigma_x^i \sigma_x^j$ is minimized if $\sigma_x^i$ and $\sigma_x^j$ 
are either both aligned or both anti-aligned with $\xi_{im}$ and $\xi_{jm}$. Thus, 
we have a two-fold degenerate ground state given by the patterns 
$\pm [{\mathrm{sign}}(\xi_{1m}),\dots,{\mathrm{sign}}(\xi_{Nm})] $. We refer to this 
scenario as the ferromagnetic side of a resonance. 

Crossing the resonance, the overall sign of the Hamiltonian changes. 
Naively, one might assume that this should not qualitatively affect the physics,  
since we have $\frac{1}{N}\sum_{i\neq j}\xi_{im} \xi_{jm} = -\frac{1}{N} \rightarrow 0$, that is, there is an equal balance between positive and negative $J_{ij}$. This expectation, 
however, turns out to be false. Recalling the relation between the Mattis model and 
number partitioning~\cite{mertens98,Mertens2003,lucas}, the antiferromagnetic model maps onto an optimization problem in which the task is to find the optimal bi-partition of a given sequence of numbers $(\xi_i)_i$, such that the cost function $E=\left(\sum_{i \in \uparrow} \xi_i - \sum_{j \in \downarrow} \xi_j\right)^2$ 
is minimized. Here, the two partitions are denoted by $\uparrow$ and $\downarrow$. For a Hamiltonian of the form
$H=\sum_{ij} \xi_{i} \sigma_x^i \xi_{j} \sigma_x^j= (\sum_i \xi_i \sigma_x^i)^2$, eigenvectors of $\sigma_x^i$ are Hamiltonian eigenstates with an energy precisely given by the cost function $E$. Thus, in the limit of just one antiferromagnetic resonance, the ground state of $H$ is exactly the configuration that minimizes the cost function. 

With this insight, the ground states of the spin model are easily found exploiting the system's parity symmetry:
For even modes, $\xi_{im}=\xi_{(N+1-i)m}$, and we simply have to choose $\langle \sigma_x^i \rangle = - \langle \sigma_x^{N+1-i} \rangle$  to minimize 
the cost function. For odd modes, $\xi_{im} = - \xi_{(N+1-i)m}$, 
and we must choose $\langle \sigma_x^i \rangle = \langle \sigma_x^{N+1-i} \rangle$. In both cases, 
this implies that we can choose half of the spins arbitrarily, leading to at least
$2^{N/2}$ ground states. 

The important observation is that an exponentially large number of ground states exists in the limit of being arbitrarily close above a resonance. This is a characteristic 
feature of spin glasses, yet it does not lead to computational hardness. In fact, as pointed out in Ref. \cite{Mertens2003}, the number partitioning problem with exponentially many perfect partitions belongs to the ``easy phase''. How to reach hard instances will be explained in the section below.

Pushing our arguments further we consider the influence of a second resonance: In between two resonances, the exponential degeneracy of the antiferromagnetic coupling on one side is lifted by the influence of the ferromagnetically coupled mode on the other side.  Interestingly, this does not lead to frustration, since even- and odd-parity modes alternate in the phonon spectrum, and the pattern favored by the ferromagnetic coupling is always contained in the ground state manifold of the antiferromagnetic coupling. Accordingly, between two modes the ground state pattern is uniquely defined by the upper mode. 

\begin{figure}[t]
\centering
\includegraphics[width=0.45\textwidth, angle=0]{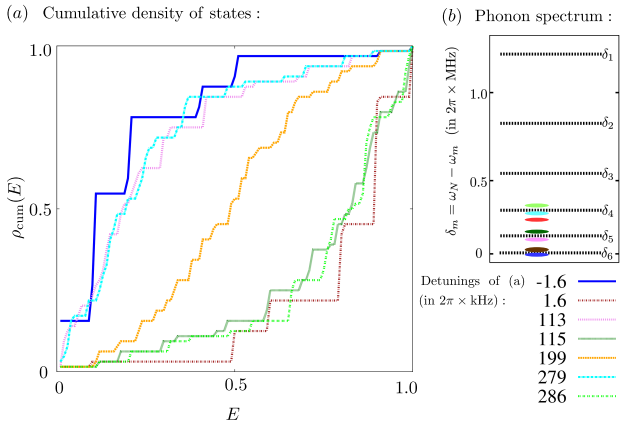}
\caption{
\label{fig:dos}
{\bf Cumulative density of states.} {\bf (a)}  For six ions and at trap frequency $\omega_z = 2\pi \times 770$ kHz, we plot the number of states (divided by the total Hilbert space dimension) below the normalized energy treshold (0: ground state energy, 1: energy of highest state), for different detunings from the center-of-mass mode. The density of states at low energies is seen to strongly increase when the detuning is slightly above a phonon resonance. In {\bf (b)}, the position of the resonances (measured from the center-of-mass mode at $\delta_m=0$) are shown.
}
\end{figure}

Beyond this two-mode approximation, we rely on numerical results. Taking into account all modes, exact diagonalization of a small system ($N\leq10$) shows that the two-mode model captures the behavior correctly: At any detuning, the degeneracy due to the nearest antiferromagnetic coupling is lifted in favor of the pattern of the next ferromagnetic coupling. In Fig.~\ref{fig:dos}(a), we plot the cumulative density of states $\rho_{\mathrm{cum}}(E)$, that is, the number of states with an energy below $E$. The corresponding phonon resonances $\omega_m$ are marked in Fig.~\ref{fig:dos}(b). The curves clearly reflect the very different behavior in the red- and blue-detuned regimes: Fig.~\ref{fig:dos} illustrates the quick increase of $\rho_{\mathrm{cum}}(E)$ at low energies, when the laser detuning is chosen on the antiferromagnetic side of a phonon resonance. In contrast, a low density of states characterizes the system on the ferromagnetic side of a resonance. In intermediate regimes, as shown for $\delta=2\pi \times 199$ kHz, the spectrum is symmetric.

A breakdown of the two-mode approximation is expected for large numbers of spins: Since the distance between neighboring resonances approximately scales with $1/N$ (at fixed trap frequencies), the influence of additional modes grows with the system size. The combined contribution of all antiferromagnetically coupled modes tries to select the fully polarized configurations as the true ground state, while all ferromagnetically coupled modes, except for the center-of-mass mode, favor fully unpolarized configurations. As a consequence, it is \textit{a priori} unclear which pattern will be selected in the presence of many modes. 

This observation is crucial from a point of view of complexity theory. In the presence of parity symmetry neither the one-mode problem (i.~e.~the number partitioning problem), nor the two-mode approximation are hard problems, as they can be solved by simple analytic arguments. However, when many modes lift the degeneracy of the exponentially large low-energy manifold in an \textit{a priori} unknown way, one faces the situation where a true but unknown ground state is separated only by a very small gap. Identifying this state then usually requires scanning an exponentially large number of low-energy states, and classical annealing algorithms can easily get stuck in a wrong minimum. Below, we discuss how a transverse magnetic field opens up a way of finding the ground state via quantum annealing. Moreover, we will discuss strategies to make also the one-mode model, i.e. the number partitioning problem, computationally complex.

\subsection*{Increasing complexity}

As discussed above, the instances of the number partitioning problem realized in the ion chain are simple to solve due to parity symmetry. This is a convenient feature when testing the correct functioning of the quantum simulation, but our goal is the implementation of computationally complex and selectable instances of the problem in the device. One strategy is the use of microtraps to hold the ions~\cite{bermudez}. The equilibrium positions of the ions can then be chosen at will, opening up the possibility to 
control the values of the $\xi_{im}$. The computational complexity of the number partitioning problem then depends on the precision with which the $\xi_{im}$ can be tuned.
If the number of digits can be scaled with the number of spins, one enters the regime where number partitioning is proven to be NP-hard~\cite{Mertens2003}. Thus, the number of digits must at least be of  order $\log_{10} N$, which poses no problem for realistic systems involving tens of ions.

Another way of enhancing complexity even within a parity-symmetric trap would be to ``deactivate'' some spins by a fast pump laser. For example, if all spins 
on the left half of the chain are forced to oscillate, $\sigma_x^j \rightarrow \sigma_x^j e^{i\omega_{\mathrm{pump}} t}$, the part of the Hamiltonian which remains time-independent poses a number partitioning problem of $N/2$ different numbers.

Another promising approach was recently suggested in Ref. \cite{haukeNP}: Operating on the antiferromagnetic side of the center-of-mass resonance, single-site addressing allows one to use the Rabi frequency for defining the instance of the number-partitioning problem. This could indeed be the step to turn the trapped ions setup into a universal number-partitioning solver, where arbitrary user-defined instances can be implemented.

If one is not interested in the number-partitioning problem itself, one might also increase the system's complexity via resonant coupling to more than one mode. Equipping the Raman laser with several 
beatnote frequencies $\omega_{\mathrm{L}}^{(\mu)}$ and Rabi frequencies $\Omega_i^{(\mu)}$, it is possible to engineer couplings of the form~\cite{korenblit-njp}:
\begin{align}
 J_{ij} \propto  \sum_{\mu=1}^{\mu_{\mathrm{max}}} \Omega_i^{(\mu)} \Omega_j^{(\mu)}  \sum_{m=1}^N  \frac{\xi_{im}\xi_{jm}}{\omega_m-\omega_{\mathrm{L}}^{(\mu)}}. 
\end{align}
With an appropriate choice of Rabi frequencies and detunings, this allows for realizing the Hopfield model, $J_{ij} \propto \sum_{\mu=1}^{\mu_{\mathrm{max}}} \xi_{im_\mu}\xi_{jm_\mu}$, 
where each coupling $\mu$ is assumed to be in resonance with one mode $m_\mu$. For ferromagnetic couplings, the low-energy states again are determined by the signs of the $\xi_{im_\mu}$, but in general the different low-energy patterns are not degenerate, and a glassy regime is expected for large $\mu_{\mathrm{max}}$ \cite{amitPRL}.

\subsection*{Quantum phases} 

\begin{figure}[t]
\centering
\includegraphics[width=0.48\textwidth]{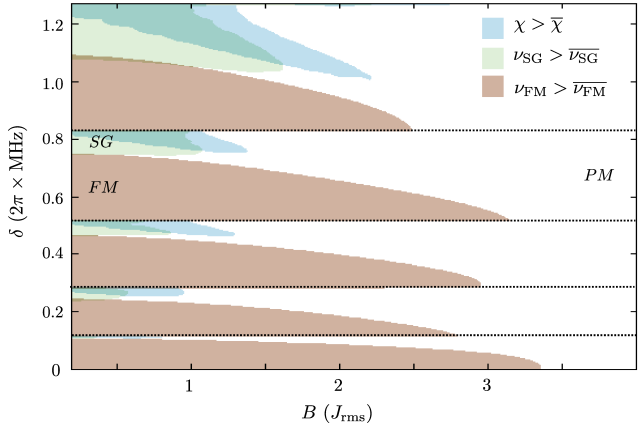}
\caption{
\label{fig:phases}
{\bf Magnetic phases.} Upon varying the laser detuning $\delta$ and the transverse magnetic field $B$, we mark, for $N=6$ ions and $\omega_z = 2\pi \times 770$ kHz, those regions in configuration space where the ferromagnetic order parameter $\nu_{\rm FM}$, the spin glass order parameter $\nu_{\rm SG}$, or the longitudinal magnetic susceptibility $\chi$ take larger than average values ($\overline{\nu_{\rm FM}}=0.13$, $\overline{\nu_{\rm SG}}=5$, $\overline{\chi}=1.6$). Below each phonon resonance (marked by the dashed horizontal lines), there is a regime where large values of $\nu_{\rm SG}$ and $\chi$ indicate spin glass behavior for sufficiently weak transverse field $B$. In order to break the $Z_2$ symmetry of the Hamiltonian, all quantities were calculated in the presence of a biasing field $\epsilon \sigma_x^1$ (with $\epsilon=-J_{\rm rms}$).
}
\end{figure}

So far, we have considered classical spin chains lacking  
any non-commuting terms in the Hamiltonian. Quantum properties come into play if we either add an additional coupling 
$\sum_{ij} \sigma_y^i\sigma_y^j$, or a transverse magnetic field:
\begin{align}
\label{HB}
H_B = \hbar B \sum_i \sigma_z^i.
\end{align}
The latter has been realized in several experiments~\cite{schaetz-natphys,kim2010,jurcevic}, and is convenient for our purposes, as the field strength 
$B$, if decaying with time, provides an annealing parameter: For large $B$, all spins are 
polarized along the $z$-direction, whereas for vanishing 
$B$ one obtains the ground state of the classical Ising chain. As 
argued above, the latter exhibits spin glass phases with an exponentially 
large low-energy subspace. Even in those cases where the true ground state 
is known theoretically, finding it experimentally remains a difficult task. 
Our system hence provides an ideal test ground for experimenting with different 
annealing strategies.

Before presenting results for the simulated quantum annealing, let us first discuss the different phases expected for the effective Hamiltonian $H_{\rm eff} = H_J + H_B + \epsilon \sigma_x^1$. The last term is a bias introduced to break the $Z_2$ symmetry. Our distinction between phases is based on certain quantities which combine thermal and quantum averages
\begin{align}
\big\langle \langle \cdot \rangle^\alpha \big\rangle_T \equiv \frac{ \sum_\lambda \bra{\lambda} \cdot \ket{\lambda}^\alpha \exp(-E_{\lambda}/k_{\rm B}T)}{\sum_\lambda \exp(-E_{\lambda}/k_{\rm B}T)},
\end{align}
with $\ket{\lambda}$ denoting Hamiltonian eigenstates at energy $E_\lambda$. For $\alpha=1$, $\langle \langle \cdot \rangle^\alpha \rangle_T$ reduces to the normal thermal average. We will use low, but non-zero temperatures $T$ of the order of the coupling constant, accounting in this way for the huge quasi-degeneracy in the glassy regime. The thermal average $\langle \cdot \rangle_T$ plays a role somewhat similar to the disorder average, as it averages over various quasi-ground states (pure thermodynamic phases). We therefore expect $\langle \langle \sigma_x^i \rangle \rangle_T$ to go to zero in the glassy phase. In contrast, a non-zero average $\langle \langle \sigma_x^i \rangle \rangle_T$ detects the ferromagnetic phase of the Mattis model, while it vanishes in the paramagnetic state. Taking its square to get rid of the sign, we obtain a global ferromagnetic order parameter by summing over all spins:
\begin{align}
 \nu_{\rm FM} = \frac{1}{N} \sum_i \langle \langle \sigma_x^i \rangle \rangle_T^2.
\end{align}
On the other hand, in the spirit of an Edwards-Anderson-like parameter, we consider thermal averages of squared quantum averages, i.e. 
\begin{align}
\nu_{\rm EA} =  \frac{1}{N} \sum_i \langle \langle \sigma_x^i \rangle^2 \rangle_T.
\end{align}
At sufficiently low temperature this average would still be zero for a paramagnetic system, but now it remains non-zero for both ferromagnetic and glassy systems. Accordingly, a parameter which is ``large'' only for glassy systems is given by the ratio $\nu_{\rm SG} = \frac{\nu_{\rm EA}}{\nu_{\rm FM}}$, used in Fig.~\ref{fig:phases} to detect glassy regions.

Thermal averages are difficult to measure, but the contained information is also present in linear response functions at zero temperature. Therefore, we have calculated the longitudinal magnetic susceptibility $\chi$, i.e. the response of the system to a small local magnetic field $h_x^j$ along the $\sigma_x$ direction (in units $J_{\rm rms}$):
\begin{align}
\label{chi}
 \chi = \frac{J_{\rm rms}}{N} \sum_{ij} \left( \frac{\partial \langle \sigma_x^i \rangle}{\partial h_x^j} \right)^2.
\end{align}
Due to the (quasi)degeneracy in the glass, one expects a huge response even from a weak field, and thus a divergent susceptibility.

We have calculated $\nu_{\rm FM}$, $\nu_{\rm SG}$, and $\chi$, for $N=6$ between $0 \leq 2\pi \times \delta \leq 8$ MHz, and $0 \leq B \leq 4 J_{\rm rms}$. For the thermal averaging, we have chosen a temperature $k_{\rm B}T = J_{\rm rms}$. The results are summarized in Fig.~\ref{fig:phases}, indicating the regions where these quantities take larger values than their configurational averages $\overline{\nu_{\rm FM}}$, $\overline{\nu_{\rm SG}}$, and $\overline{\chi}$, defined as $\overline{f} \equiv \int {\rm d}B \int {\rm d}\delta f(B,\delta)/(B_{\rm max}\delta_{\rm max})$. In this way, we identify and distinguish ferromagnetic behavior below, and glassy behavior above each resonance. Regions of large susceptibility $\chi$ overlap with regions of large $\nu_{\rm SG}$, attaining numerical values which are three orders of magnitude larger than the corresponding averages. For sufficiently strong field $B$, in contrast, none of these parameters is large, indicating paramagnetic behavior.

Note that the existence of the purported glassy phase in the quantum case is an open problem. We provide here the evidence only for small systems, since it is numerically feasible and corresponds directly to current or near-future experiments. If we increase the complexity of our system by resonant coupling to many phonon modes, as discussed in the last paragraph of the previous subsection, the glassy behavior will result from the interplay of contributions of many modes – similarly as in the Hopfield model with Hebbian rule and random memory patterns. Here the beautiful results by Strack and Sachdev \cite{sachdev2011} -- the  “quantum” analog of the Amit~{\it et al.} \cite{Amit1} machinery -- can be applied directly to obtain the phase diagram for large $N$. If, however,  we increase the complexity by random positioning of the ion traps, then the resonance condition will pick up the contribution from one dominant (random) mode, and the Hebbian picture will apply.

\subsection*{Simulated quantum annealing} 

We will now turn to the more realistic description of the system in terms of a time-dependent Dicke model, described by the Hamiltonian $H_0$ in Eq.~(\ref{H0}) with an additional transverse field $H_B(t)$ from Eq.~(\ref{HB}). We assume an exponential annealing protocol $B(t) = B_{\rm max} \exp(-t/\tau)$.  Again we apply a bias field $h_{\rm bias} = \hbar \epsilon \sigma_x^i$, lifting the $Z_2$ degeneracy of the classical ground state. We study the exact time evolution  under the Hamiltonian $H(t)=H_0(t)+H_B(t)+h_{\rm bias}$, using a Krylov subspace method \cite{park}, and truncating the phonon number to a maximum of two phonons per mode.

Initially, the system is cooled to the motional ground state, and spins are polarized along $\sigma_z$. Choosing $B_{\rm max} \gg \epsilon, J_{\rm rms}$, this configuration is close to the ground state of the effective model $H_J+H_B+h_{\rm bias}$ at $t=0$. If the decay of $B(t)$ is slow enough, and if the entanglement between spins and phonons  remains sufficiently low, the system stays close to the ground state for all times, and finally reaches the ground state of $H_J$. 

\begin{figure}[t]
\centering
\includegraphics[width=0.498\textwidth]{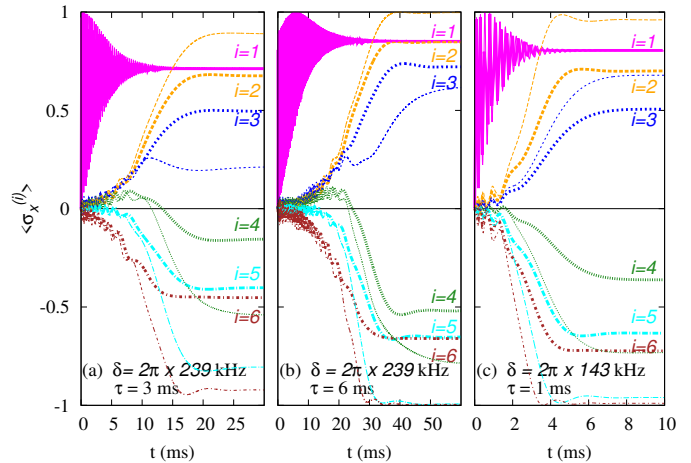}
\caption{
\label{fig:anneal}
{\bf Simulated quantum annealing in the closed system.} Unitary time evolution of the full system ($N=6$) for different laser detunings $\delta$ between the 4th and 5th resonance. Thick lines show the result of the exact evolution, while the thin lines have been obtained from the semi-classical approximation. In all cases, the desired pattern $(+++---)$ can be read out after sufficiently long annealing times. While in {\bf (a,b)}, we operate at the onset of glassiness, $\delta=2\pi \times 239$ kHz, the panel (c) considers a ferromagnetic instance, $\delta=2\pi \times 143$ kHz. The annealing protocol in (a) is defined by 
$B_{\rm max} = 50 J_{\rm rms}$ and $\tau=3$ ms. In (b), the same instance is solved with higher fidelity by choosing $B_{\rm max} = 80 J_{\rm rms}$ and $\tau=6$ ms. Fast annealing, with $B_{\rm max} = 50 J_{\rm rms}$ and $\tau=1$ ms, is possible in the ferromagnetic instance in (c). In all simulations, we have chosen $\Omega=2\pi \times 50$ kHz, $\omega_{\rm recoil} = 2\pi\times 15$ kHz, and $\epsilon=-10$ kHz.
}
\end{figure}

We have simulated this process for $N=6$ ions, as shown in Fig.~\ref{fig:anneal}. As a result of the annealing, we are not interested in the final quantum state, but only in the signs of $\langle \sigma_x^i \rangle$, which fully determine the system in the classical configuration. This provides some robustness. We find that for a successful annealing procedure, yielding the correct sign for all $i$, the number of phonons produced during the evolution should not be larger than 1. At fixed detuning, we can reduce the number of phonons by decreasing the Rabi frequency, at the expense of increasing time scales. As a realistic choice \cite{jurcevic,richerme}, we demand that annealing is achieved within tens of miliseconds. 

Fig.~\ref{fig:anneal}(a) shows that one can operate at a detuning $\delta = 2\pi\times 239$ kHz, that is, at the onset of a glassy phase according to Fig.~\ref{fig:phases}.  The mode vector which selects the ground state is $\boldsymbol\xi_5=(0.61,0.34,0.11,-0.11,-0.34,-0.61)$, and the corresponding pattern can be read out after an annealing time $t\geq 15$ ms. On the other hand, even for long times, $\langle \sigma_x^{i=4} \rangle$ saturates only at $-0.15$, which is far from the classical value $-1$. As shown in Fig.~\ref{fig:anneal}(b), a slower annealing protocol leads to more robust results ($|\langle \sigma_x^i \rangle|>0.52 \ \ \forall \ i$). In Fig.~\ref{fig:anneal}(c), a much simpler instance in the ferromagnetic regime is considered. Good results ($|\langle \sigma_x^i \rangle| > 0.36 \ \  \forall \ i$) can then be obtained within only a few ms.

In addition, dephasing due to instabilities of applied fields and spontaneous emission processesundesired processes disturb the dynamics of the spins. In Ref.~\cite{haukeNP} a master equation was derived that takes into account such noisy environment. To study the evolution of the system in this open scenario we have applied the Monte Carlo wave-function method \cite{molmer}. As quantum jump operators are hermitean, $\sigma_x^i$ for dephasing and $\sigma_z^i$ for spontaneous emission, the evolution remains unitary, but is randomly interrupted by quantum jumps. Each jump has equal probability $\Gamma$, and the average number of jumps within the annealing time $T$ is given by $\overline{n_{\rm jumps}} = 2 N \Gamma T$, which we chose close to 1.

\begin{figure}[t]
\centering
\includegraphics[width=0.498\textwidth]{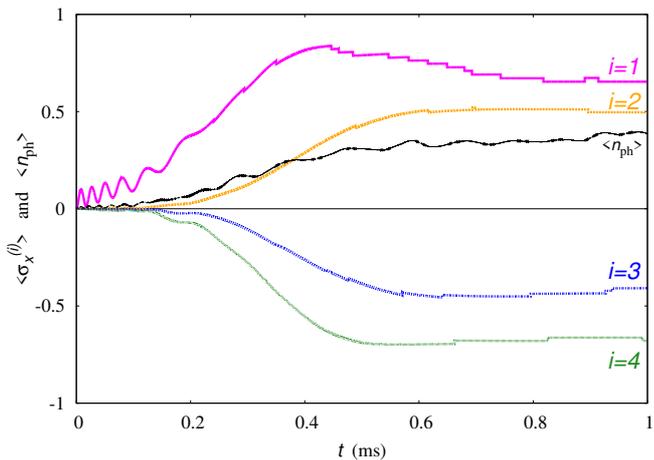}
\caption{
\label{fig:mcwf}
{\bf Simulated quantum annealing in the open system.} Using a Monte Carlo wave-function description averaged over 100 runs, we simulate the annealing process with $N=4$ ions in the presence of dephasing and spontaneous emission. With a noise rate $\Gamma = 0.03 J_{\rm rms}\approx 120$ Hz and a total annealing time $T=1$ ms, we have on average one jump per run, $\overline{n_{\rm jumps}} = 0.96$. We plot expectation values of the spins and total number of phonons, produced by the coupling, as a function of time. Here, we have operated in the ferromagnetic regime between the second and third phonon resonance, $\delta=2\pi\times 159$ kHz. Choosing $\Omega=2\pi \times 50$ kHz, $\omega_{\rm recoil} = 2\pi\times 15$ kHz, $\omega_z= 2\pi \times 876$ kHz, $B_{\rm max} = 50 J_{\rm rms} \approx 200$ kHz, and $\epsilon=-10$ kHz, we are able to perform very fast annealing, $\tau=0.1$ ms, with high fidelity. After a time $T= 1$ ms, the unitary time evolution has converged to $(\langle \sigma_x^1 \rangle, \langle \sigma_x^2 \rangle, \langle \sigma_x^3 \rangle, \langle \sigma_x^4 \rangle) = (0.97, 0.70,-0.65,-0.85)$, correctly reproducing the classical pattern $(++--)$.
}
\end{figure}

Since a faithful description requires statistics over many runs, we restrict ourselves to a small system, $N=4$, with short annealing times. In a sample of 100 runs, we noted 94 jumps (42 $\sigma_x$ and 52 $\sigma_z$ jumps). In 39 runs, no jump occured. Amongst the 61 runs in which at least one jump occured, 26 runs still produced the correct sign for all spin averages $\langle \sigma_x^i \rangle$. The full time evolution, averaged over all runs, is shown in Fig.~\ref{fig:mcwf}. On average, the final result is $(0.65,0.50,-0.41,-0.68)$, that is, our annealing with noise still produces the correct answer, but with lower fidelity.

Whether an individual jump harms the evolution crucially depends on the time at which it occurs: While a spin flip ($\sigma_z$ noise) is harmless in the beginning of the annealing, a dephasing event ($\sigma_x$ noise) at an early stage of the evolution leads to wrong results. Oppositely, at the end of the annealing procedure, dephasing becomes harmless while spontaneous emission falsifies the result. An optimal annealing protocol has to balance the effect of different noise sources against non-adiabatic effects in the unitary evolution.

\subsection*{Scalability}
Above we have demonstrated the feasibility of the proposed quantum annealing scheme in small systems. The usefulness of the approach, however, depends crucially on its behavior upon increasing the system size. While the exact treatment of the dynamics becomes intractable for longer chains, an efficient description can be derived from the Heisenberg equations:
\begin{align}
 \label{eom}
& i\hbar \frac{ \rm d}{{\rm d}t} \langle a_m \rangle = \langle \left[ a_m,H(t)\right] \rangle, \\
& i\hbar \frac{ \rm d}{{\rm d}t} \langle \sigma_\alpha^i \rangle = \langle \left[ \sigma_\alpha^i,H(t) \right] \rangle, \nonumber
\end{align}
with $H=H_0(t) + H_B(t) + h_{\rm bias}$. To solve this set of $5N$ first-order differential equations, we make a semi-classical approximation $\langle a_m \sigma_x^i \rangle \approx  \langle a_m \rangle \langle \sigma_x^i \rangle$, and then proceed numerically using a fourth order Runge-Kutta algorithm. The semi-classical approximation is justified as long as the system remains close to the phonon vacuum. A direct comparison with exact results for six ions (see Fig.  \ref{fig:anneal}) shows that the semiclassical approach, while slightly overestimating fidelities, accurately reproduces all relevant time scales.

This approach allows us to extend our simulations up to $N=22$ ions, at trap frequency $\omega_z = 2\pi \times 270$ kHz. We operate between the first and second resonance where the level spacing is largest, at a beatnote frequency $\omega_{\rm L}=\omega_1+0.2(\omega_2-\omega_1)$, that is with a fixed relative detuning between the two modes. This choice corresponds to $\delta=2\pi \times 1.2$ MHz in Fig.~\ref{fig:phases}, characterized as a glassy instance of the system. 

\begin{figure}[t]
\centering
\includegraphics[width=0.498\textwidth]{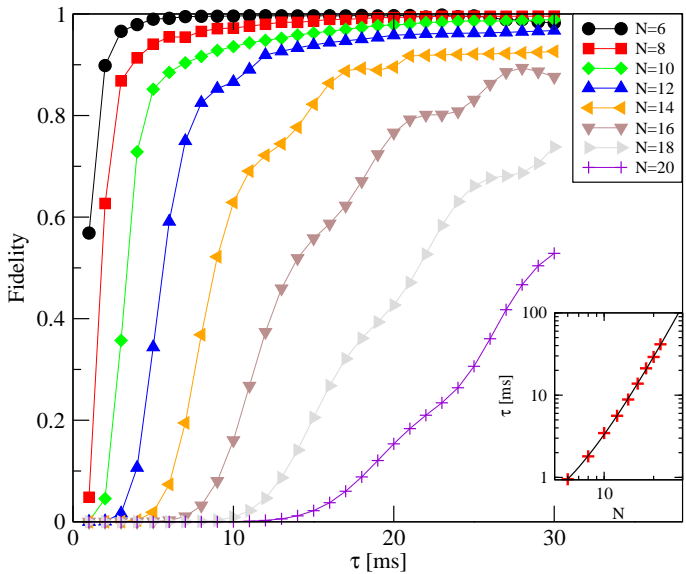}
\caption{
\label{fig:fidelity}
{\bf Fidelity estimates of the quantum annealing and scaling of annealing times.} We solve the equations of motion, Eq. (\ref{eom}), for a glassy instance at different system sizes $N$, and plot the fidelity of the outcome as a function of the annealing time $\tau$. In the inset, we investigate the scaling behavior by plotting (in double-logarithmic scale) the value of $\tau$ which is needed for a fidelity $F=0.5$ as a function of $N$. A fourth-order polynomial fit agrees very well with the data (black dashed line). The fit parameters, as defined by Eq. (\ref{fit}), are $\tau_0 = (90 \pm 40)$ ms and $\gamma=12.0 \pm 1.2$. For all calculations, we have chosen a beatnote frequency between the two lowest resonances, $\omega_{\rm L}=0.8\omega_1+0.2\omega_2$, a trap frequency $\omega_z=2\pi\times 270$ kHz, and a bias potential $\epsilon = -1$ kHz. The initial value of the transverse field was $B_{\rm max}=10$ kHz.}
\end{figure}

Our aim is to find the relation of annealing time, measured by the decay parameter $\tau$, and systems size while the fidelity $F$ is kept constant. For a practical definition we demand that $F$ is zero when the annealing fails, that is when the sign of the spin averages $\langle \sigma_x^i\rangle$ does not agree with the classical target state for all $i$. If the annealing finds the correct signs, the robustness still depends on the absolute values of the spin averages. The fidelity is then defined as the smallest absolute value, $F={\rm min}_i |\langle \sigma_x^i \rangle|$. Our results are summarized in Fig.~\ref{fig:fidelity}: 
Firstly, this figure shows that for all sizes $N\leq 20$ large fidelities $F\geq 0.5$ can be produced within experimentally feasible time scales, $\tau\leq 30$ ms. 
Secondly, the time scale $\tau$ needed for a fidelity $F=0.5$ fits well to a fourth-order polynomial in $N$ (with subleading terms of the order $\exp(1/N)$):
\begin{align} 
\label{fit}
\tau(N) =  N^4 \tau_0 \exp(\gamma/N),
\end{align}
with $\tau_0$ and $\gamma$ being free fit parameters. Although the sample of 22 ions is too small to draw strong conclusions, it is noteworthy that the polynomial fit is more accurate than an exponential one. This suggests that the proposed quantum simulation is indeed an efficient way of solving a complex computational problem. One should also keep in mind that our estimates, based on a semi-classical approximation, neglect certain quantum fluctuations which could further speed-up the annealing process.

To study the scaling of dissipative effects, we have extendend the Monte Carlo wave function approach to larger systems, which is feasible if the phonon dynamics is neglected. The unitary part of the evolution is then described by the effective Ising Hamiltonian $H_{\rm eff}=H_J+H_B(t)+h_{\rm bias}$. The dissipative part consists of random quantum jumps described by $\sigma_x^i$ and $\sigma_z^i$. The results for a glassy instance ($\delta=2\pi \times 198$ kHz at $\omega_z = 2\pi\times 700$ kHz) are summarized in Table \ref{table} for $N=4, 6, 8$. The noise rate is chosen such that on average one quantum jump occurs in the system with four ions, while accordingly the system with eight ions suffers on average from two such events. In all cases, the annealing produces the correct pattern, $F>0$. As expected, $F$ decreases for larger systems, but fortunately rather slowly (from $F=0.25$ at $N=4$ to $F=0.16$ at $N=8$). If the total amount of noise is kept constant, i.e. $\Gamma \propto 1/N$, the annealing is found to profit from larger system sizes, since a quantum jump at spin $i$ is unlikely to affect the sign of $\langle \sigma_x^j \rangle $ for $j\neq i$. We note that the spin values produced by the Monte Carlo wave function method cannot be described by a normal distribution. Importantly, the peak of each distribution, roughly coinciding with its median, is barely affected by the noise. Thus, larger fidelities can be obtained from the median rather than from the arithmetic mean of $\langle \sigma_x^i \rangle$.

\begin{table*}
\begin{ruledtabular}
\begin{tabular}{lcrr}
& closed system:& $(1.0, 0.97, -0.39, -0.42 )$ \\
$N=4$ & open system, mean values: & $(    0.79,    0.70,   -0.25,   -0.32)$\\
& open system, median values:& $(1.0, 0.97,-0.39,-0.42$)\\
 \hline
& closed system: &    $(1.0,    0.99,    0.76,   -0.40,   -0.64,   -0.66$)\\
 $N=6$ & open system, mean values.& $( 0.82,    0.68,    0.55,   -0.23,   -0.43,   -0.46)$ \\
  & open system, median values:& $( 1.0,    0.98,    0.76,   -0.39,   -0.64,   -0.66)$ \\
 \hline
 & closed system:& $( 1.0,    0.99,    0.96,    0.58,  -0.31,   -0.70,   -0.75,   -0.77$)\\
 $N=8$ & open system, mean values:& $(0.83,    0.69,    0.69,    0.40,  -0.16,   -0.43,   -0.55,   -0.53)$ \\
  & open system, median values:& $(1.0,    0.98,    0.94,    0.58,   -0.30,   -0.69,   -0.75,   -0.76)$ \\
 \end{tabular}
\end{ruledtabular}
\caption{\label{table}
{\bf Annealing in effective spin model with and without noise.} We perform quantum annealing ($\tau=3$ ms and $T=30$ ms) for a glassy instance ($\delta=2\pi \times 198$ kHz at $\omega_z = 2\pi\times 700$ kHz, $\epsilon=-1$ kHz) at different system sizes $N$, using an effective spin model description. The results $(\langle \sigma_x^1 \rangle, \dots, \langle \sigma_x^N \rangle)$, are shown for the closed-system dynamics and for a noisy system, with mean and median over a sample of 2000 runs. The average number of noisy event scales with the system size, and is adjusted to $4/N$. In all cases, the signs of $\langle \sigma_x^i\rangle$ reproduce correctly the mode pattern, and the fidelity decreases with the system size. In contrast to the arithmetic mean values, the median values in the noisy sample are barely affected by the noise. 
}
\end{table*}

\subsection*{Spin pattern in the quantum Mattis model}
The quantum annealing discussed above exploits quantum effects in order to extract information encoded in the classical model. Now we search for information which is encoded in the quantum, but not in the classical model. Therefore, we focus on the ferromagnetic Mattis model, which in the classical case keeps a binary memory of a spin pattern, that is, of $N$ bits. Our considerations can also be generalized to the Hopfield model~\cite{hopfield}, which memorizes multiple patterns. We will show how quantum effects can increase the amount of information encoded by these models.

\begin{figure}[t]
\centering
\includegraphics[width=0.45\textwidth, angle=0]{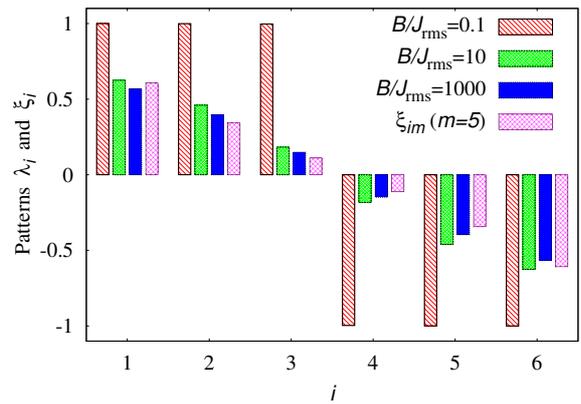}
\caption{
\label{fig:magma}
{\bf From binary to real-valued patterns.} The spin expectation values $\lambda_i$ approach the real values of the mode vector $\xi_{im}$ when a sufficiently strong transverse magnetic field $B$ is present. The shown data was obtained from the effective spin Hamiltonian for $N=6$ ions in the ferromagnetic regime, $\delta=2\pi\times 143$ kHz. Deviations from the equality $\lambda_i=\xi_{im}$ are smaller than 0.04.
}
\end{figure}

Recall that in the classical case, the spin pattern was defined by a resonant mode in terms of the sign of each component. In the quantum case, however, one cannot simply replace classical spins by quantum averages, ${\mathrm{sign}}(\langle \sigma_x^i \rangle)$. Even in a weak transverse field $B$, this quantity vanishes due to the $Z_2$ symmetry, $\sigma_x \rightarrow -\sigma_x$. Instead, the pattern is reflected by $\lambda_i = \bra{\Psi_1} \sigma_x^i \ket{\Psi_2}$, where $\ket{\Psi_1}$ and $\ket{\Psi_2}$ are the ground and first excited state. For small $B$, we find numerically ${\mathrm{sign}}(\lambda_i) = {\mathrm{sign}}(\xi_{im})$. For large $B$, the stronger relation $\lambda_i = \xi_{im}$ holds approximately, see Fig.~\ref{fig:magma}. Thus, the former binary memory has become real-valued.

To show this behavior, we note that for strong $B$, the ground state is 
fully polarized along $z$, and the first excited state is restricted to the $N$-dimensional subspace with one spin flipped, 
that is, $S_z = \sum_i \sigma_z^i = N-2$. Within this subspace the Hamiltonian 
$H_J$ is given by an $N\times N$ matrix approximately proportional to 
$\tilde J_{ij} = -\xi_{im} \xi_{jm}$ for $i\neq j$, and $\tilde J_{ii}={\rm constant}$. Here we neglect all but the $m$-th mode close to resonance.

It is easy to see that the vector $\boldsymbol\xi_m$ is a ground state of the matrix $-\xi_{im}\xi_{jm}$, which differs from $\tilde J_{ij}$ only by the diagonal elements, which approach unity for large $N$. The first excited state reads $\ket{\Psi_2} = \sum_{i=1}^N \xi_{im} \ket{i}$, where $\ket{i}$ denotes the state in which spin $i$ is flipped relatively to all others (in the $\sigma_z$ basis). This shows that $\lambda_i \approx \xi_{im}$, and the small deviations decrease quickly with $N$. 

Measuring $\lambda_i$ experimentally is possible by full state tomography. The absolute value of $\lambda_i$ can be obtained via a simple $\sigma_z^i$ measurement. In the limit of strong $B$-fields we have $\lambda_i = [(1-\langle \sigma_z^i \rangle)/2]^{1/2}$.

%
%
Many applications are known for the classical spin system with couplings defined by spin patterns, reaching from pattern recognition and associative memory in the Hopfield model~\cite{hopfield} to noise-free coding~\cite{nishimori,sourlas}. Our analysis suggests that patterns given by real numbers could replace patterns of binary variables by exploiting the quantum character of the spins. 

\section*{\large             Discussion                 }

In summary, our work demonstrates the occurrence of Mattis glass behavior in spin chains of trapped ions, if the detuning of the spin-phonon coupling is chosen between two resonances. In these regimes, the effective spin system has an exponentially large number of low-energy states, and finding its ground state corresponds to solving a number-partitioning problem. This establishes a direct connection between the properties of a physical system and the solution of a potentially NP-hard problem of computer science. Given the state-of-art in experiments with trapped ions, the physical implementation is feasible: In comparison to previous experiments with trapped ions \cite{schaetz-natphys,monroe-spinspin,kim2010,britton2012,jurcevic}, only the detuning of the spin-phonon coupling needs to be adjusted. Differently from other approaches to spin glass physics, our scheme does not require any disorder. In its  most natural implementation, parity symmetry allows one to analytically determine the ground state. Different ways to break this symmetry can be implemented to increase the complexity of the problem.

The ion chain then becomes an ideal test ground for applying quantum simulation strategies to solve computationally complex problems. By applying a transverse field to the ions, quantum annealing from a paramagnet to the glassy ground state is possible. 
The ionic system may  be used to benchmark quantum annealing, which has become a subject of very lively and controversial debate since the launch of the D-Wave computers \cite{troyer2014,katzgraber}. Exact calculations for small systems ($N=6$) and approximative calculations for larger system ($N=22$) demonstrate the feasibility of the proposed quantum annealing, and suggest a polynomial scaling of the annealing time. Accordingly, this approach may offer the sought-after quantum speed-up. In view of sizes of 30 and more ions already trapped in recent experiments (cf. Ref. \cite{ramm}), a realization of our proposal could not only confirm our semi-classical results, but also go beyond the sizes considered here.

Finally, resonant coupling to multiple modes opens an avenue to neural network models, where a finite number of patterns is memorized by couplings to different phonon modes. Quantum features can increase the memory of such networks from binary to real-valued numbers. It will be subject of future studies to work out the possible benefits which quantum neural networks may establish for information processing purposes.

\textit{ Acknowledgements.}
We thank A. Ac\'in, R. Augusiak, A. Bermudez, J. Tura, P. Rotondo, L. Santos, and 
P. Wittek for discussions. We acknowledge financial support from EU grants OSYRIS 
(ERC-2013-AdG Grant No. 339106), SIQS (FP7-ICT-2011-9 No. 600645), 
QUIC (H2020-FETPROACT-2014 No. 641122), EQuaM (FP7/2007-2013 Grant No. 323714), 
Spanish MINECO grants (FOQUS FIS2013-46768-P, FIS2014-54672-P), Severo Ochoa grant SEV-2015-0522), Generalitat de Catalunya (2014 SGR 401, 2014 SGR 874 and SGR 875), the Maria de Maeztu grant (MDM-2014-0369), and Fundaci\'o Cellex. B.J-D.\ is supported by the Ramon y Cajal program, and C.G. by MPQ-ICFO, ICFOnest+ (FP7-PEOPLE-2013-COFUND), and by the European
Union's Marie Sklodowska-Curie Individual Fellowships (IF-EF) programme (GA 700140).

\textit{Contributions.}
All authors contributed to the conceptual design of the quantum simulation, discussion and interpretation of results, and the writing of the manuscript. Exact numerical calculations were performed by TG. Semiclassical calculations were done by BJD and DR.

\textit{Competing financial interests.}
The authors declare no competing financial interests.


\end{document}